\documentclass[11pt,aps,prd,preprint,pacs,nofootinbib,superscriptaddress]{revtex4-1}
\usepackage{amsmath}
\usepackage{epsf}
\usepackage{graphicx}  
\usepackage{epstopdf}
\usepackage{dcolumn}   
\usepackage{bm}        
\usepackage{color}
\usepackage[english]{babel}
\usepackage{wasysym}
\begin{document}
\title{The entropy of Hawking radiation and the generalized second law}
\author{Diego Pav\'{o}n\footnote{E-mail: diego.pavon@uab.es}}
\affiliation{Departamento de F\'{\i}sica, Facultad de Ciencias,
Universidad Aut\'{o}noma de Barcelona, 08193 Bellaterra
(Barcelona), Spain}
\begin{abstract}
\noindent We derive an approximate expression for the entropy of
Hawking radiation filling a spherical box in stable thermodynamic
equilibrium with the Schwarzschild black hole that produced the
said radiation. The Bekenstein entropy  bound is satisfied but the
generalized second law might not be always guaranteed. We briefly
discuss the possible origin of this unexpected result.
\end{abstract}
\maketitle
\section{Introduction}
\noindent The stability of thermodynamic equilibrium between a
Schwarzschild black hole and its own radiation was considered long
ago \cite{diego-werner}. A spherical, uncharged, classical black
hole of mass $M$ can remain in  equilibrium with its emitted
radiation, enclosed in a spherical box of perfecting reflecting
thin walls of radius $R$ $\, (R>2M)$, provided that total energy
inside the box ($M + E_{rad}$) ($E_{rad}$ being the energy of
Hawking's radiation in the box) has a minimum. For this minimum to
exist the total energy must be  distributed between both
components in such a way that the ratio $M/E_{rad}$ is not greater
than a precise value that varies for every pair $(M, R)$.
Actually, no stable equilibrium is possible if $E_{rad}> M/4$.
This confirms the finding of Hawking \cite{hawking1976} who (for
mathematical simplicity) assumed the radiation to be purely
thermal.

\noindent It is expedient to briefly recall the meaning of
thermodynamic stable equilibrium in a macroscopic system.
First, the temperature (as well other intensive thermodynamic
parameters) must fulfill Tolman's law \cite{Tolman1934} across the
system and the macroscopic parameters (energy volume, number of
particles, etc) are to stay constant. Second, the spontaneous
microscopic fluctuations  of temperature, pressure, etc, around
the equilibrium state die away very quickly. In consequence, the
macroscopic flows of energy, momentum, particles, etc, remain zero. By
contrast, when the equilibrium is not stable the aforementioned
fluctuations grow fastly and dissipative flows inevitably arise.
In general, the latter take the system to a new stable equilibrium
state of a larger entropy than the original.

\noindent Our aim is threefold: $(i)$ to obtain a well behaved,
entropy function for the Hawking radiation enclosed in the
aforesaid box when stable equilibrium between the Schwarzschild
hole and radiation prevails, $(ii)$ to see whether the Bekenstein
bound \cite{bekensteinconstraint} on the ratio entropy/energy is
satisfied, and $(iii)$ to check whether the generalized second law
(GSL) holds in the evaporation process of the black hole. This law
(one should be mindful that actually rather than a law it is a
reasonable conjecture) asserts that the Bekenstein-Hawking entropy
of a black hole plus the entropy of its surroundings cannot
decrease in time.

\noindent We shall use Page's approximate stress-energy tensor of
Hawking radiation \cite{page1982} alongside Euler's thermodynamic
equation. A very reasonable expression for the entropy of Hawking
radiation follows (Section I). In section II we check  the
Bekenstein bound, and in section III we explore whether the GSL is
fulfilled. As it turns out, it may be violated when the size of
the system is not much larger  than the  wavelength of the typical
Hawking radiation mode. In last Section we summarizes and discuss
our findings.

\noindent We do not impose the evaporation process of the black
hole to be unitary. As is well known, this assumption was adopted
in several studies (see e.g., \cite{page2013, almheiri2021} and
references therein) of the evolution of the entropy of Hawking
radiation in order to overcome the famous information loss
paradox. Here we do not deal with this highly debatable issue given
the lack of consensus on it.

\noindent  We choose units such that $\hbar = c = G = k_{B} =1$.


\section{Hawking radiation}
\noindent Here we derive an approximate expression for the entropy
of Hawking radiation filling a spherical thin box of radius $R$,
of perfectly reflecting walls, in thermodynamic equilibrium with
the black hole that produced the radiation. The formula we arrive
to is not an exact one because: $(i)$ we start from the expression
for a conformally invariant scalar field in the Hartle-Hawking
thermal state  around a Schwarzschild black hole provided by Page,
\cite{page1982} which, in spite of  being very good, is only
approximate. $(ii)$ We ignore the vacuum polarization induced by
the gravitational field of the black hole. Although the
polarization dies away quickly with curvature (the latter is given
by $48 M^{2} r^{-6}$) is not negligible near the horizon. $(iii)$
We also ignore the effect of the vacuum polarization of the wall
of the box, and ($iv$) the back reaction of the radiation on the
metric.

\noindent We begin by recalling the metric of a Schwarzschild black hole in spherical coordinates,
$ ds^{2} = - f(r) \, dt^{2} \, + \, f^{-1}(r) \, dr^{2} \, + \, r^{2} (d \theta^{2} \, +
\, \sin^{2} \theta \, d \phi^{2})$, and Tolman's law, $T = f^{-1/2}(r) \, T_{bh}$, where
$f(r) =  1- 2M/r$ and $T_{bh} = (8 \pi M)^{-1}$.

\noindent  Page's stress-energy tensor  reads  \cite{page1982},
\begin {equation}
\overline{T}^{\mu}_{\, \nu} = \frac{\pi^{2}}{90}\left(\frac{1}{8
\pi M}\right)^{4}\,  \left\{\frac{1 - (4-\frac{6M}{r})^{2}
\left(\frac{2M}{r} \right)^{6}}{(1 - \frac{2M}{r})^{2}} \,
(\delta^{\mu}_{\, \nu} - 4 \delta^{\mu}_{\, 0} \delta^{0}_{\,
\nu}) \, + \ 24 \left(\frac{2M}{r}\right)^{6} \, (3
\delta^{\mu}_{\, 0} \delta^{0}_{\mu}+ \delta^{\mu}_{\, 1}
\delta^{1}_{\, \mu})\right\}. \label{page-tensor}
\end{equation}
\noindent This quantity is conserved and has the right trace. From
it the energy density ($-\overline{T}^{0}_{\, 0}$) and pressure
($\overline{T}^{k}_{\, k}/3$)
\begin{equation}
\rho = \frac{\pi^{2}}{30} \, \left(\frac{1}{8 \pi M}\right)^{4}\,
\left\{\frac{1}{(1-\frac{2M}{r})^{2}} \, \left[1-
\left(4-\frac{6M}{r}\right)^{2} \,
\left(\frac{2M}{r}\right)^{6}\right] - 24
\left(\frac{2M}{r}\right)^{6}\right\},
 \label{rho2}
\end{equation}
\begin{equation}
P = \frac{\pi^{2}}{90} \, \left(\frac{1}{8 \pi
M}\right)^{4}\,\left\{\frac{1}{(1-\frac{2M}{r})^{2}} \, \left[1-
\left(4-\frac{6M}{r}\right)^{2} \, \left(\frac{2M}{r}\right)^{6}
\right]\, + \, 8 \left(\frac{2M}{r}\right)^{6}\right\}, \label{P2}
\end{equation}
\noindent of Hawking radiation follow.

\noindent When $r \gg M$ the equation of state is that of thermal
radiation in flat space-time, $P = \rho/3$, as it should.

\noindent The entropy density is given by Euler's equation, $
s_{rad} = T^{-1} (\rho + P)$ \cite{Callen1960} with the chemical
potential set to zero because this quantity vanishes for thermal
radiation. Thus,
\begin{equation}
s_{rad} = \frac{2 \pi^{2}}{45}
\left(\frac{1}{8 \pi M}\right)^{3} \left(1 -
\frac{2M}{r}\right)^{1/2} \left\{\frac{1- \left(4-
\frac{6M}{r}\right)^{2} \left(\frac{2M}{r}\right)^{6}}{\left(1 -
\frac{2M}{r}\right)^2} - 64 \left(\frac{2M}{r}
\right)^{6}\right\},
\label{srad2}
\end{equation}
\noindent it goes to zero as  $r \rightarrow 2M$ and tends
asymptotically to the thermal flat space-time value $(2\pi^{2}/45)
T^{3}$ when $ r \gg M$.

\noindent The radiation entropy contained in a box of radius R
is given by $S_{rad} = \int_{2M}^{R}{4\pi r^{2} f(r)^{-1/2}\,
s_{rad} \, dr}$.
\noindent Therefore,
\[
S_{rad} = \frac{2 \pi^{2}}{45} \left(\frac{1}{8 \pi M}\right)^{3}
\left\{32 \pi M^{3} \left[\frac{1}{3}\left(\frac{R}{2M}-1
\right)^{3}- 2 \left(\frac{R}{2M}-1 \right)^{2}  + 6
\left(\frac{R}{2M}-1 \right) + 4 \ln \left(\frac{R}{2M}\right)
\right. \right.
\]
\begin{equation}
\left. \left. +1 - \left(\frac{2M}{R}\right)^{2} + 3
\left(\left(\frac{2M}{R}\right)^{3}-1\right) + 5\left(1 -
\frac{2M}{R}\right) +
\frac{64}{3}\left(\left(\frac{2M}{R}\right)^{3}-1\right)
\right]\right\}. \label{Srad2}
\end{equation}

\noindent  This expression vanishes for $R \rightarrow 2M$, it is bounded from
below (at $(R/2M)_{min} \simeq 1.742$ it has a local minimum), and
diverges when $M \rightarrow 0$. It is negative in the range $\,1
\leq R/2M \lesssim 4.912$, which should not  surprise us
because the energy density (\ref{rho2}) is negative when $r$ is
not much larger than $2M$. Thereby the GSL can be violated in that
interval. As expected, for $R \gg 2M$ the thermal flat space-time
expression $\, S_{rad}\mid_{R \gg 2M} \rightarrow \frac{2
\pi^{2}}{45}\, \frac{4 \pi}{3} \, R^{3} T_{bh}^{3}$ is recovered.

\noindent It is seen that the entropy of Hawking
radiation depends just on the ratio $R/M$ and not on $R$
or $M$ separately. This is only natural after our using of
Euler equation and integrating.

\noindent Reasonably, Euler's equation (and thereby,  Eqs.
(\ref{srad2}) and (\ref{Srad2}))  holds when the size of the box
is at least three times larger than the wavelength of the typical
Hawking radiation mode. The latter is of the order of the black
hole radius \cite{birrell-davies}, $r_{s} = 2M$. Note also that
for $r \geq 3 r_{s}$ the curvature is bounded from above by $1/972
M^{4}$; whence for $M \geq 1$ it results negligible.

\begin{figure}[!htb]
 \begin{center}
    \begin{tabular}{c}
     \resizebox{70mm}{!}{\includegraphics{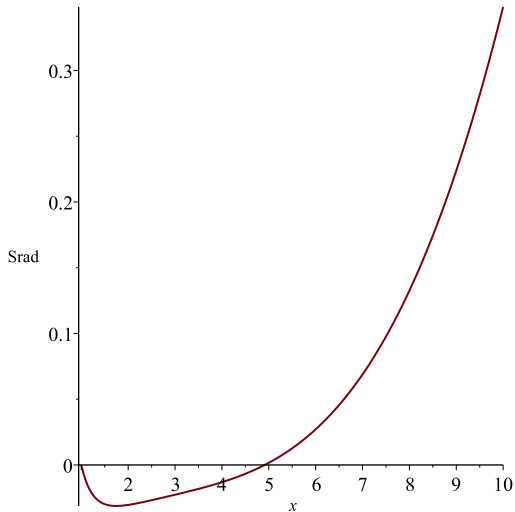}}\\
    \end{tabular}
    \caption{The entropy of Hawking radiation, Eq. (\ref{Srad2a}), appears  to be negative in the
    interval $1 \leq x \lesssim 4.912$. However, see the text.}
    \label{fig:4-dim-Srad}
 \end{center}
 \end{figure}

\noindent For later convenience we rewrite (\ref{Srad2}) in terms
of the dimensionless parameter $x= R/2M$,
\newpage
\[
S_{rad} = \frac{1}{360} \, \left[\frac{1}{3}(x-1)^{3} - 2(x-1)^{2}
+ 6(x-1) + 4 \ln x \right.
\]
\begin{equation}
\left.  - 5(x^{-1}- 1) - (x^{-2} -1) +
\frac{73}{3}(x^{-3}-1)\right].
 \label{Srad2a}
\end{equation}
\noindent A plot of this expression is shown in Fig.
\ref{fig:4-dim-Srad}. The fact that $S_{rad}$ is negative in some
interval may be an un-physical consequence of: $(i)$ $R$ being of
the order of $M$ in that interval, $(ii)$ having ignored the back
reaction of the radiation on the metric and the vacuum
polarization effects of the wall of the box.

\section{Bekenstein's entropy bound}
\noindent In 1981 Bekenstein  argued that the ratio between the
entropy and total energy, $E$, of a system cannot be arbitrarily
large but subject to the constraint \cite{bekensteinconstraint}
\begin{equation}
\frac{S}{E}\leq 2 \pi R \, ,
\label{bekenstein-constraint1}
\end{equation}
where $R$ is the radius of the circumference that circumscribes
the system. This conjecture proved controversial, see e.g.
\cite{page2005,page2008} and references therein; nevertheless, it
bears some interest and we will explore it in the case at hand.

\begin{figure}[!htb]
 \begin{center}
    \begin{tabular}{c}
     \resizebox{70mm}{!}{\includegraphics{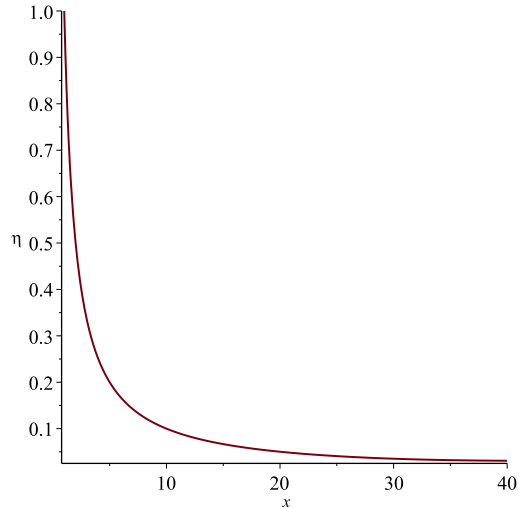}}\\
    \end{tabular}    \caption{$\eta(200,x)$ {\it vs} $x$. $\eta$ is upper-bounded by unity.}
    \label{fig:eta2}
 \end{center}
\end{figure}

\noindent Upon defining $\eta = S/(2 \pi R E)$ with $S = 4 \pi
M^{2} + S_{rad}$, where $S_{rad}$ is given by (\ref{Srad2a}) and
$E= M + E_{rad}$ with
\begin{equation}
E_{rad} = \int_{2M}^{R}{4 \pi r^{2} \rho \, dr} = \frac{1}{1920
\pi R}\left[\frac{x^{4}}{3}+ x^{3}+ 3x^{2}+ 4 x \ln x - \frac{22}
{3} x - 5 - 3 x^{-1}+ 11 x^{-2}\right],
\label{eq:erad2}
\end{equation}
we have
\begin{equation}
\eta = \frac{\pi R^2 x^{-2} + \frac{1}{360} \, \left[\frac{1}{3}(x-1)^{3} - 2(x-1)^{2}
+ 6(x-1) + 4 \ln (x) - 5(x^{-1}- 1) - (x^{-2} -1) + \frac{73}{3}(x^{-3}-1)\right]}
{\pi R^2 x^{-1} + \frac{1}{960}\left[\frac{x^{4}}{3}+ x^{3}+ 3x^{2}+ 4 x \ln x - \frac{22}
{3} x - 5 - 3 x^{-1} + 11 x^{-2} \right]}.
\end{equation}
\noindent From last expression it is seen that $0 <\eta \leq 1$;
more precisely,  $\eta(x \rightarrow 1) = 1$ and $\eta(x>1) \sim
x^{-1}$. Altogether, the entropy's Bekenstein bound is fulfilled.
Figure \ref{fig:eta2} illustrates this for the case $R =200$.

\noindent It is worth noting that because the entropy of any
system is maximum at equilibrium, this bound is also satisfied
when the black hole and radiation are out of mutual thermal
equilibrium.
\section{The Generalized Second Law}
\noindent As noted above, according to the GSL, the entropy of a
black hole plus the entropy of its surroundings cannot decrease
with time. On the one hand this seems rather natural and, to the
best of our knowledge, no covincing counterexample has been
provided thus far. On the other hand, it has never been really
proved. Here we examine whether our system (a Schawarschild black
hole plus Hawking radiation in a spherical box) obeys this law.

\noindent Notice that Page's stress-energy tensor
(\ref{page-tensor}) was obtained assuming a static space-time
whereby $M$ was necessarily constant. However, if $M$ is large
enough, the evaporation rate will be very small, namely: $\dot{M}
\simeq -\alpha(M)\, M^{-2}$ \cite{page1976}, with $\alpha(M)$ a
small and nearly constant quantity for most of the black hole
life-time. Therefore, in this situation temporal derivatives are
admissible and equations (\ref{dStotal2}) and (\ref{lambda2}),
below,  may be considered valid  so long as $M \gg 1$.

\noindent The temporal derivative of the total entropy can be
written as,
\begin{equation}
\dot{S} = \dot{S}_{bh} + \dot{S}_{rad} = \lambda(R,x) \,
\mid\dot{M}\mid
\label{dStotal2}
\end{equation}
\noindent with
\begin{equation}
\lambda(R,x) = - \frac{4 \pi R}{x} \, + \, \frac{1}{180 R}
\left[x^{4} - 6 x^{3}+11 x^{2}+ 4 x +5 +2 x^{-1} - 73
x^{-2}\right]. \label{lambda2}
\end{equation}
\begin{figure}[!htb]
 \begin{center}
    \begin{tabular}{c}
    \resizebox{70mm}{!}{\includegraphics{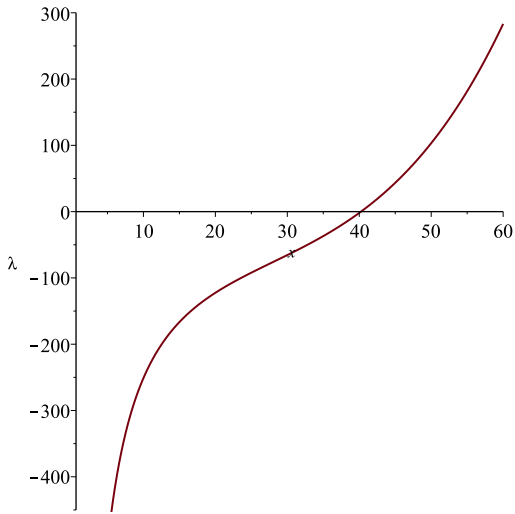}}\\
    \end{tabular}
    \caption{$\lambda(R,x)$ {\it vs} $x$, for $R = 200$. The curve
    crosses the horizontal axis at $x_{0} \simeq 40.24$. Beyond that point the GSL is satisfied. It
    is not  in the interval $x_{min} < x < x_{0}$ with $x_{min}
    \simeq38.558$. See the text.}
    \label{fig:dS2}
 \end{center}
\end{figure}

\begin{figure}[!htb]
 \begin{center}
    \begin{tabular}{c}
     \resizebox{70mm}{!}{\includegraphics{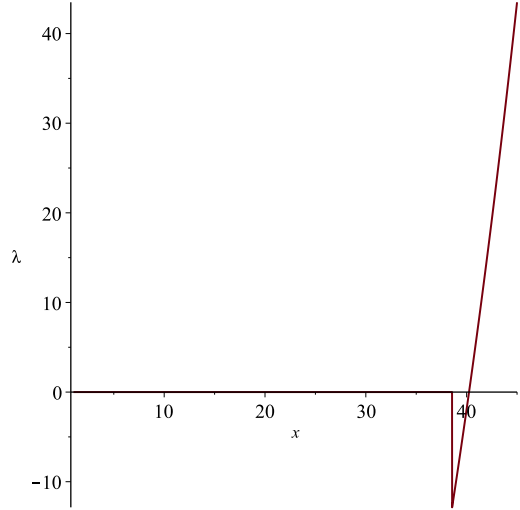}}\\
    \end{tabular}
    \caption{In the interval $1< x \leq x_{min}$ ($x_{min} \simeq 38.558$, with $R = 200$)
    stable thermodynamic equilibrium  between the black
    hole and Hawking radiation holds, hence $\dot{M} = \dot{S} = 0$.
    Beyond $x_{min}$ the equilibrium is unstable, therefore between
    $x_{min}$ and $x_{0}$ one has $\dot{M} \neq 0$ and $\dot{S} < 0$  as $\lambda$ is negative.
    For $x \geq x_{0} \simeq 40.24$$, \lambda$ is positive and so is $\dot{S}$. The ascending line,
    starting at $x_{min}$, corresponds to the graph of $\lambda (200, x \geq x_{min})$.}
    \label{fig:4-dim-piecewise}
 \end{center}
\end{figure}

\noindent This equation is somewhat misleading as it seems to
imply that the GSL is violated ($\dot{S} < 0$) in the interval $1
< x \leq x_{0}$, where $x_{0}$ is implicitly defined by setting
$\lambda(R,x)$ to zero, see Fig. \ref{fig:dS2}. In reality, it
does appear violated only when, alongside $\lambda$ being negative
the system is out of stable equilibrium. As shown in Ref.
\cite{diego-werner}, for $R = 200$ in the interval $1 < x \leq
x_{min} \simeq 38.558$ (see table 2 in the said reference) the
black hole is in stable thermodynamic equilibrium with the
radiation. As a consequence, $\dot{M} = 0$ and so $\dot{S}  = 0$
as well. It is just in the subinterval  $x_{min} < x <x_{0}$ (with
$x_{0} \simeq 40.24$) that $\lambda <0$ and, simultaneously, the
system is in an unstable thermodynamic equilibrium state whence
$\dot{S} <0$ and the GSL is violated. Naturally, when $x > x_{0}$
the GSL is satisfied since the equilibrium is unstable (whence
$\dot{M} \neq 0$) and $\lambda >0$. Fig. \ref{fig:4-dim-piecewise}
illustrates this situation. As numerical calculation confirms,
this pattern holds true regardless the radius of the box.
Therefore though, formally, $\dot{S} < 0$ in the interval $(1,
x_{min}]$, the entropy will remain constant in that interval
(because stable thermodynamic equilibrium prevails in it). Then,
$\, S$ will decrease and later on (from $x_{0}$ onwards) increase.
Consider for instance the case shown in Fig.
\ref{fig:4-dim-piecewise}. Imagine that initially $x = 39$ (a
value in the subinterval $(x_{min}, x_{0})$); since there is no
stable equilibrium, the black hole will loose mass to Hawking
radiation continuously. This alongside the size of the box being a
constant implies that $x$ will increase unbounded, and sooner or
later $\dot{S}$ will become positive (right after $\lambda$ had
crossed the horizontal axis). Thus even if the GSL is not
fulfilled initially, it will eventually.
\begin{figure}[!htb]
 \begin{center}
    \begin{tabular}{c}
     \resizebox{70mm}{!}{\includegraphics{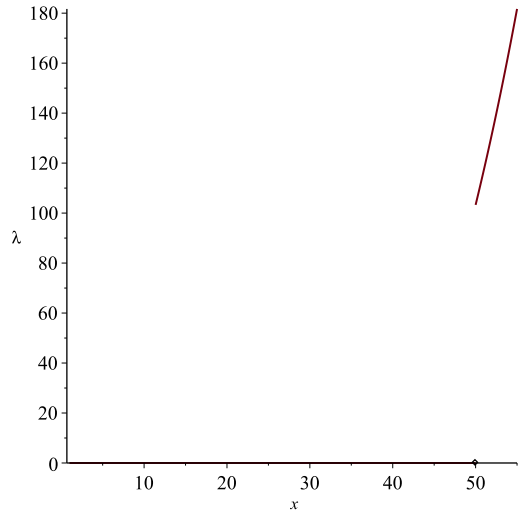}}\\
    \end{tabular}
    \caption{In drawing the plot, for $R = 200$,  it was artificially assumed that $x_{min} = 50$,
    larger than $x_{0} \simeq 40.24$. As a result the $\lambda$ curve never lies below the horizontal
    axis but it experiences an un-physical discontinuity, namely $\Delta \lambda \simeq 103.28$
    in our units.}
    \label{fig:4-dim-discont}
 \end{center}
\end{figure}

\noindent Clearly if $x_{min}$ could be greater than $x_{0}$, the
region where $\lambda$ is negative would be non-existent and
$\dot{S}$ would never become negative. However, as Fig.
\ref{fig:4-dim-discont} depicts, this is not possible, i.e., the
inequality $x_{min} > x_{0}$ is not to be satisfied ever. If it
were, the $\lambda$ curve would undergo an un-physical
discontinuity, $\Delta \lambda(R, x)$, at $x = x_{min}$. It would
mean that right after the loss of stable thermal equilibrium
between the black hole and Hawking radiation, the initial
$\dot{S}$ instead of being close to zero (though positive), would
be $\lambda(R, x_{min}) \mid \dot{M} \mid$. As it can be
numerically verified, this result does not depend on the value of
$R$. Therefore, the GSL won't be violated whenever $x_{min}$
coincides with $x_{0}$.

\subsection{ A Gedanken experiment}
\noindent  Consider a black hole of mass $\, M_{1}$ surrounded by
spherical box, of radius $\, R_{1 }$, of perfectly reflecting thin
walls centered at the black hole, and further away a similar box
of radius $\, R_{2} > R_{1}$. Also assume that the black hole is
in stable thermodynamic equilibrium with the Hawking radiation
filling the inner box. Between both boxes there is a vacuum. Then,
imagine that the smaller box is gently removed whence the Hawking
radiation (originally filling just the smaller box) expands and
fills the larger box while it  gets colder and falls out of
equilibrium with the hole. By the Le Ch\^{a}thelier-Braun
principle \cite{Callen1960} the black hole reacts  and gets hotter
by emitting Hawking radiation. Now, two distinct possibilities,
depending on whether the total heat capacity of the system is
negative or positive, arise. In the first case, a new stable
equilibrium, at a higher temperature, will be achieved. In the
second, no stable equilibrium is feasible \cite{diego-werner,
hawking1976, diego-miguel1983} and by a continuous evaporation
process the black hole will shrink more and more while the
radiation temperature rises enormously (the details of this
process, especially his last stages, in which the black hole may
even disappear altogether, are not well-known; actually, it is
most likely a subject of quantum gravity). In any case, the
Bekenstein-Hawking entropy of the black hole will decrease as
$T_{bh}^{-2}$ whereas the radiation entropy will rise
approximately as $T_{bh}^{3}$. Consequently, the final entropy
will be larger than the entropy of the system before removing the
inner box. We shall focus on the first possibility, namely, that
of a final state of thermodynamic stable equilibrium.

\noindent In this thought  experiment the total energy is
conserved (i.e., $M_{1} + E_{rad1} = M_{2} + E_{rad2}$). If we fix
the values of $\, R_{1}, x_{1}$ and $R_{2}$, then, $x_{2}$ will be
set by the energy conservation equation (a constraint equation).
Once $x_{2}$ is found, we can determine whether the total entropy
($S = S_{bh} + S_{rad}$) augments or decreases in the process $\,
(M_{1}, R_{1}) \Rightarrow (M_{2}, R_{2}$). Since the latter is
irreversible, one should  expect the final entropy to be larger
than the initial one. But this is not ensured beforehand and it
should be checked.
\begin{figure}[!htb]
 \begin{center}
    \begin{tabular}{c}
     \resizebox{70mm}{!}{\includegraphics{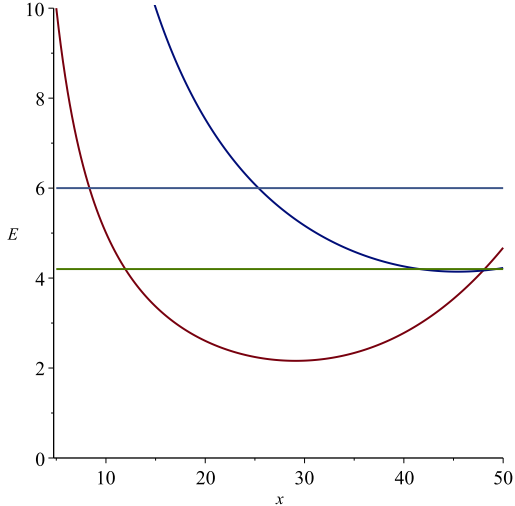}}\\
    \end{tabular}    \caption{Total energy {\it vs} $x$ assuming the radius of the inner and outer boxes being $100$ and $300$,
    respectively (from the left curve to the right one). The horizontal straight lines correspond to $E = 4.2$ and $E = 6$
    in our units. The lower one first intersects the left and right curves at $x\simeq 11.946$ and  $x \simeq 41.608$, respectively.
    The corresponding intersection points of the upper one are  $\, x \simeq 8.339$ and $\, x \simeq 25.365$.}
    \label{fig:4-dim-energy2}
 \end{center}
\end{figure}
Fig. \ref{fig:4-dim-energy2} can be interpreted as follows. Let us
focus first on the lower straight line. In this case, the system
has a total energy $E= 4.2$ and it is in stable thermodynamic
equilibrium. Then, the inner box, of radius $R_{1}= 100$, is
removed and the radiation expands up to the larger box, of radius
$R_{2} = 300$. The system attains a new stable equilibrium but at
a higher temperature. In this process the quantity $x$ increases
from about $11.946$ up to about $41.608$ and the black hole mass
goes down by a factor close to $0.86$. Next we focus on the upper
straight line. In this instance the total energy is $8$;
everything parallels the previous case except that now $x_{1}
\simeq 8.339$, $x_{2} \simeq 25.365$, and the mass of the black
hole decreases by a factor of about $0.98$.

\noindent We are now poised to calculate the variation of entropy
in both cases. Clearly
\begin{equation}
\Delta S = \pi \left(\frac{R_{2}^{2}}{x_{2}^{2}} - \frac{R_{1}^{2}}{x_{1}^{2}}\right) + S_{rad2} - S_{rad1},
\label{4-dim-DeltaS}
\end{equation}
where $S_{rad1}$ denotes $\, S_{rad}(M_{1}, R_{1})$, and
analogously so does $S_{rad2}$.

\noindent Upon inserting the numerical values quoted above ($R_{1}
= 100$, $R_{2} = 300$, $x_{1} = 11.946$ and $x_{2} = 41.608$) for
the case  $E = 4.2$  in last expression, it yields $\Delta S
\simeq - 4.02$; this is to say, the GSL is not satisfied.

\noindent For the $E =6$ case, we obtain $\, \Delta S \simeq
-4.536$. Thus, the GSL is violated also in this instance.

\noindent At any rate, inspection of (\ref{4-dim-DeltaS}) shows that for
$x_{2}$ large enough, keeping $R_{1}$ and $x_{1}$ fixed, the GSL
holds true.

\noindent One may think that the GSL could be easier analyzed by
plotting the total entropy ($S = 4 \pi M^{2} + S_{rad}$) as a
function of $x$. However, one should be mindful of the two only
possibilities, namely: $(i)$ The system, black hole plus Hawking
radiation, is in state of stable thermodynamic equilibrium. In
such a case $M$ as well $S_{rad}$ remain constant. $(ii)$ The
system is out of equilibrium. In this second case, the expression
for $S_{rad}$, Eq. (\ref{Srad2a}), does not apply.

\section{Summary and Discussion}
\noindent By means  of Page stress-energy tensor
(\ref{page-tensor}) and Euler's thermodynamic equation we derived
an approximate expression, Eq. (\ref{Srad2}), for the entropy of
Hawking radiation  filling a thin spherical box of perfectly
reflecting walls. The said expression assumes thermal stable
equilibrium between the black hole and radiation.  It has some
nice features; among other things,  it solely depends on the ratio
between the radius of the box and the black hole mass but not on any
of these two quantities by themselves.

\noindent While the Bekenstein bound \cite{bekensteinconstraint}
is fulfilled, our results suggest that the GSL might be violated
not only when $R$ is of the order of $M$ but also when it is
larger since the temporal derivative of
the total entropy decreases with time. Also a Gedanken experiment
that compares the total entropy after and before removing a
intermediate box around the black hole shows violation of
the GSL. The latter is recovered when the bigger box
is much greater than the intermediate one.

\noindent The violation of the GSL is surprising. However, at this stage,
it would be premature to claim that it is the case. This result
may arise from the fact that Page's stress-energy tensor, despite
being a very good approximation, is not exact, especially near the black
hole horizon where the energy density and entropy of the radiation
become negatives. Further, we have ignored the back reaction of
the radiation on the metric and the effect of the vacuum
polarization of the wall of the box. Further, we might have overlooked
some other  source of entropy.  Nevertheless, we believe our
study points out the convenience to consider the subject more
carefully. Actually, its solution may not be round the corner and,
possibly, we are to wait for a reliable theory of quantum gravity
to settle the matter.

\noindent Nonetheless, even if the GSL  is not fulfilled in some
phase of the evaporation it will be globally
respected. Indeed, consider the entropy of a cloud of $N$
particles, $S_{cloud} \sim N$ (see e.g. {\cite{frautschi1982}),
just before undergoing a full gravitational collapse. The
Bekenstein-Hawking entropy of the ensuing black hole will be
$S_{bh} \sim N^{2}$ and the GSL satisfied. However, the partial or
total evaporation of this object results into a cloud of particles
whose entropy will be again given by the above expression,
$S_{new-cloud} \sim N_{new-cloud}$ (in general both numbers will
differ, and in the last phases of the evaporation $N_{new}$ will
attain a  much bigger value than $N$). In any case, even though
the GSL might be  transcended in the evaporation process, the total
process,
\newline \centerline{Initial cloud $\rightarrow$ Black Hole
$\rightarrow$ Evaporation $\rightarrow$ Final cloud,}
\newline may well comply with the GSL on account of the inequality
$N_{new-cloud} > N$. If the GSL is fulfilled in the evaporation,
then we would also have $\, N_{new-cloud} > N^{2}$. Obviously, this
is just a hand-waving argument whence it is desirable to verify
its outcome  by a more rigorous study.

\noindent Finally, it would be worthwhile to establish a relation
between our expression of Hawking entropy and the expression for
the entanglement entropy derived by Funai and Sugawara
\cite{funai2021} on the principle of local field theory and the
assumption of entanglement between the radiation and the black
hole interior. However, this approach presents the drawback that
for the process to be unitary the singularity at the center of the
Schwarzschild hole must be replaced by a solid core. Maybe this
difficulty will disappear when a better understanding of the
process itself is achieved.
\section*{Acknowledgments}
\noindent Thanks are due to Narayan Banerjee, Don Page and Bin
Wang for useful comments on an earlier version of this manuscript.

\end{document}